\documentstyle{article}
\title{Scale invariant Euclidean  field theory in any dimension}
\author{ Z. Haba\\Institute of Theoretical Physics, University of Wroclaw,
\\50-204 Wroclaw, Plac Maxa Borna 9,Poland\\e-mail:zhab@ift.uni.wroc.pl}
\date{}
\begin{document}
\maketitle
\baselineskip 24pt
\begin{abstract}
We discuss $D$-dimensional scalar field interacting with  a
scale invariant random  metric which is either a Gaussian field or a square
of a Gaussian field.
 The metric depends  on $d$ dimensional
coordinates (where $d<D$).  By a projection to a lower
dimensional subspace  we obtain
a scale invariant
non-Gaussian model of Euclidean quantum field theory
in $D-d$ or $d$ dimensions.

\end{abstract}

\section{Introduction}
We consider a new method of a construction of Euclidean fields. A
scalar field in $D$ dimensions is interacting with a metric
depending on $d$ dimensional coordinates. An averaging over the
metric and a projection of the scalar field to an $s$
dimensional  subspace leads to a scalar field which is Euclidean
invariant in $R^{s}$ (we consider $s=D-d$ and $s=d$). If the
metric field is scale invariant with a scaling dimension
$2\gamma$ then the scalar field is also scale invariant with a
scaling dimension depending on $\gamma$. We discuss two models
for the random metric. In the first model we consider a square of
a Gaussian random field. We are unable to derive an upper bound
for correlation functions in this model. Then, we consider a
metric which is Gaussian. We obtain scale invariant correlation
functions with explicit upper and lower bounds. Our primary
interest in this class of models \cite{brze}-\cite{PLB} comes
from quantum gravity. However, the method may be useful for a
construction of relativistic  quantum fields (although at the
moment we are unable to prove the crucial Osterwalder-Schrader
positivity \cite{ost}). The model can be interesting for
statistical physics as a continuum version of spin glass models
\cite{parisi}. The lattice version of our model describes spins
with a random
 coupling  between them which is either Gaussian (then we have
 a mixing between
ferromagnetic and antiferromagnetic couplings)
or a square of the Gaussian field. A calculation of
the average over the random coupling can be done
explicitly. As a result we obtain models with  many
interacting spins in contradistinction to the conventional
models  based on bilinear spin-spin interactions.

\section{$D$-dimensional scalar fields}
We consider a complex scalar matter field $\Phi$ in $D$ dimensions
interacting with gravitons varying only on a $d$-dimensional
submanifold. We split the coordinates as
  $x=(X,{\bf x})$ with ${\bf x}\in R^{d}$.
Without a self-interaction the $\Phi\Phi^{*}$ correlation function
is equal to an average ($W(g)$ is the gravitational action)
\begin{equation}
\int {\cal D}g\exp\left(-\frac{1}{\hbar}W\left(g\right)\right)
{\cal A}^{-1}(x,y)
\end{equation}
over the gravitational field $g$ of the Green's
function  ${\cal A}^{-1}(x,y)$ of the operator
\begin{equation}
{\cal A}=\frac{1}{2}
\sum_{\mu=0,\nu=0}^{D-d-1}g^{\mu\nu}({\bf x})\partial_{\mu}\partial_{\nu}+
\frac{1}{2}\sum_{k=D-d}^{D-1}\partial_{k}^{2}
\end{equation}
In order to calculate the average (1) we repeat some steps of
refs.\cite{brze}-\cite{PLB}. We represent the Green's function by
means of the proper time method
\begin{equation}
{\cal A}^{-1}(x,y)=\int_{0}^{\infty}d\tau\left(\exp\left(\tau {\cal A}\right)
\right)(x,y)
\end{equation}
For a calculation of   $\left(\exp\left(\tau {\cal A}\right)
\right)(x,y)   $ we apply the functional integral
\begin{equation}
\begin{array}{l}
K_{\tau}(x,y)=\left(\exp\left(\tau {\cal A}\right)
\right)(x,y)=\int {\cal D}x\exp(-\frac{1}{2}\int \frac{d{\bf x}}{dt}
  \frac{d{\bf x}}{dt}-\frac{1}{2}\int g^{\mu\nu}({\bf x})\frac{dX_{\mu}}{dt}
  \frac{dX_{\nu}}{dt})
 \cr
 \delta\left(x\left(0\right)-x\right)
  \delta\left(x\left(\tau\right)-y\right)
  \end{array}
  \end{equation}
In the functional integral (4) we make a change of variables ($x
\rightarrow b$) determined by Stratonovitch stochastic
differential equations \cite{ike}
\begin{equation}
dx^{\Omega}(s)=e_{A}^{\Omega}\left(
x\left(s\right)\right)db^{A}(s)
\end{equation}
where for $\Omega=0,1,....,D-d-1$
\begin{displaymath}
e^{\mu}_{a}e^{\nu}_{a}=g^{\mu\nu}
\end{displaymath}
and $e^{\Omega}_{A}=\delta^{\Omega}_{A}$ if $\Omega>D-d-1$.

After such a change of variables the functional integral in
eq.(4) becomes Gaussian.  In fact, this is the standard Wiener
integral and $b_{A}(t)$ for each $A$ are independent Brownian
motions
\begin{displaymath}
E[b_{A}(t)b_{C}(s)]=\delta_{AC}\min(s,t)
\end{displaymath}
The solution $q_{\tau}$ of eq.(5) consists of two vectors $({\bf
Q},{\bf q})$ where
\begin{equation}
{\bf q}(\tau,{\bf x})={\bf x}+ {\bf b}(\tau)
\end{equation}
and ${\bf Q}$ has the components (for $\mu=0,...,D-d-1$)
\begin{equation}
Q^{\mu}(\tau,{\bf X})=X^{\mu}+\int_{0}^{\tau}
e_{a}^{\mu}\left({\bf q}\left(s,{\bf x}\right)\right)dB^{a}(s)
\end{equation}
The kernel is
\begin{displaymath}
\begin{array}{l}
K_{\tau}(x,y)=E[\delta(y-q_{\tau}(x))] \cr = E[\delta({\bf y}
-{\bf x}-{\bf b}(\tau)) \prod_{\mu}\delta\left(Y_{\mu}-Q_{\mu}
\left(\tau,X\right)\right)]
\end{array}
\end{displaymath}
Using eq.(7) and the Fourier representation of the $\delta$-function
we write the kernel $K_{\tau}$ in the form
\begin{equation}
\begin{array}{l}
K_{\tau}(x,y)=(2\pi)^{-D+d}\int d{\bf P}
 \exp\left(i{\bf P}\left({\bf
Y}-{\bf X}\right)\right)
 \cr E[\delta\left(
{\bf y}-{\bf x}-{\bf
b}\left(\tau\right)\right)
\exp\left(-i\int P_{\mu}e^{\mu}_{a}\left ({\bf
q}\left(s,{\bf x}\right)\right)dB^{a}\left(s\right)\right)]
\end{array}
\end{equation}
We may choose a Gaussian  field as a model for the tetrad
(as we did in ref.\cite{brze})
\begin{equation}
\langle e^{\mu}_{a}({\bf x})e^{\nu}_{b}({\bf y})\rangle=
\Gamma^{\mu\nu}_{ab}({\bf x}-{\bf y})=
\alpha^{\mu\nu}_{ab}\vert {\bf x}-{\bf y}\vert^{-2\gamma}
\end{equation}
where $\alpha $ is a scale invariant tensor. Then
\begin{equation}
\begin{array}{l}
\langle K_{\tau}(x,y)\rangle=(2\pi)^{-D+d}\int d{\bf P}
\exp
\left( i{\bf P}  \left({\bf Y}    -{\bf X}\right) \right)

 \cr E[\delta \left({\bf y}-{\bf
x}    -\sqrt{\tau}{\bf b}\left(1\right)\right)
\exp\left(-\tau^{1-\gamma} P_{\mu}
P_{\nu}\int_{0}^{1}dB^{a}\left(s\right)\int_{0}^{s}dB^{c}
\left(s^{\prime}\right)\Gamma^{\mu\nu}_{ac}\left ({\bf
b}\left(s\right)-{\bf b}\left(s^{\prime}\right)\right)
\right)]
\end{array}
\end{equation}
where we have changed the time $s\rightarrow \tau s$ ,
used the equivalence $b(\tau s)=\sqrt{\tau}b(s) $
and the scale invariant form of the two-point function (9).
Moreover, we renormalized the kernel $K_{\tau}$ removing
from it the term (see \cite{brze}\cite{luk})
\begin{displaymath}
 \exp(-\frac{1}{2}\tau\Gamma^{\mu\nu}_{aa}(0)P_{\mu}P_{\nu})
 \end{displaymath}
 It can be seen that this procedure is equivalent to the
 normal ordering of the metric as a square of the tetrad
 \begin{equation}
g^{\mu\nu}(x)=e^{\mu}_{a}(x)e^{\nu}_{a}(x)\rightarrow
       :e^{\mu}_{a}(x)e^{\nu}_{a}(x):=
       e^{\mu}_{a}(x)e^{\nu}_{a}(x)
      -\langle e^{\mu}_{a}(x)e^{\nu}_{a}(x)\rangle
 \end{equation}
 We can prove that the double stochastic integral in eq.(10)
 is a finite square integrable random variable
 if $2\gamma<1$. However, it remains unclear whether
 the momentum integral in eq.(10) is finite.

We can work without the stochastic integrals (8) if we explicitly
integrate over $B$. The random variables ${\bf b}$ and $B^{a}$
are independent. Hence, using the formula \cite{ike}
\begin{displaymath}
E[\exp i\int f_{a}({\bf q})dB^{a}]=
E[\exp(-\frac{1}{2}\int f_{a}f_{a}ds)]
\end{displaymath}
we can rewrite eq.(8) solely in terms of the metric tensor
\begin{equation}
\begin{array}{l}
K_{\tau}(x,y)=(2\pi)^{-D+d}\int d{\bf P}
         \exp\left(i{\bf P}\left({\bf Y}-{\bf
X}\right)\right)
 \cr E[\delta\left(
{\bf y}-{\bf x}-{\bf b}\left(\tau\right)\right)
\exp\left(-\frac{1}{2}\int_{0}^{\tau} P_{\mu}g^{\mu\nu}\left
({\bf q}\left(s,{\bf x}\right)\right)P_{\nu}ds\right)]
\end{array}
\end{equation}
Let ${\cal J}$ be the characteristic function of $g^{\mu\nu}$
\begin{equation}
{\cal J}(h)=\langle
\exp\left(-\frac{1}{2}g\left(h\right)\right)\rangle
\end{equation}
Then, the mean value of the kernel (12) can be expressed in the form
\begin{equation}
\begin{array}{l}
\langle K_{\tau}(x,y)\rangle=(2\pi)^{-D+d}\int d{\bf P}

\exp\left(i{\bf P}\left({\bf Y}-{\bf X}\right)\right) \cr
E[\delta\left( {\bf y}-{\bf x}-{\bf b}\left(\tau\right)\right)
 {\cal J}(h)]
\end{array}
\end{equation}
where
$ g(h)=\int d{\bf z}g^{\mu\nu}({\bf z})h_{\mu\nu}({\bf z})$
and
\begin{displaymath}
h_{\mu\nu}({\bf z})=P_{\mu}P_{\nu}\int_{0}^{\tau}\delta\left({\bf z}-{\bf x}-
{\bf b}\left(s\right)\right)ds
\end{displaymath}
If $e^{\mu}_{a}$ is Gaussian then ${\cal J}$ can be calculated
explicitly
\begin{equation}
\langle \exp\left(-\frac{1}{2}g\left(h\right)\right)\rangle
=\det\left(1+h\Gamma\right)^{-\frac{1}{2}}
\end{equation}
where on the r.h.s. the renormalization of the determinant
(through  a multiplication by
$\exp\left(\frac{1}{2}Tr\left(\Gamma h\right)\right)$
defining $\det_{2}$, see \cite{det}) is equivalent to the normal ordering (11) (and
subsequently to the renormalization of  the kernel (10)).

We consider next another (simpler) model where the metric
is Gaussian with
two-point correlations
\begin{equation}
\langle g^{\mu\nu}({\bf x})g^{\sigma\rho}({\bf y})\rangle
=- D^{\mu\nu;\sigma\rho}({\bf x}-{\bf y})=-
C^{\mu\nu;\sigma\rho}
 \vert {\bf  y}-{\bf  x}\vert^{-4\gamma}
\end{equation}
where $C$ (a scale invariant operator) must be positive definite
if the momentum integrals in the final formula are to exist. This
requirement is not satisfied in a linearized Einstein gravity
\cite{wein} ( e.g.,in the transverse-traceless gauge
$p_{\Omega}p_{\Gamma}g^{\Omega\Gamma}(x)$  would be zero in a
covariant $D$-dimensional gravity;however our gravity is
$d$-dimensional). The conformally flat metric
$C^{\mu\nu;\sigma\rho}=\delta^{\mu\nu}\delta^{\sigma\rho}$ would
be a satisfactory model for our purposes .

The average  over $g$ in eq.(12) can be  calculated
\begin{equation}
\begin{array}{l}
\langle K_{\tau}(x,y)\rangle=(2\pi)^{-D+d}\int d{\bf P}
 \exp\left( i{\bf P}  \left({\bf Y}    -{\bf X}\right)\right)
 \cr E[\delta \left({\bf y}-{\bf
x}    -\sqrt{\tau}{\bf b}\left(1\right)\right)\cr
\exp\left(-\frac{1}{4}\tau^{2-2\gamma}\int_{0}^{1} P_{\mu}P_{\sigma}
P_{\nu}P_{\rho}D^{\mu\nu;\sigma\rho}\left ({\bf
b}\left(s\right)-{\bf b}\left(s^{\prime}\right)\right)
dsds^{\prime}\right)]
\end{array}
\end{equation}
(as in eq.(10) we have changed the time $s\rightarrow \tau s$). By
a scaling  of momenta we can bring the propagator of eq.(3) to
the form
\begin{equation}
\langle {\cal A}^{-1}(x,y)\rangle=\int_{0}^{\infty}d\tau
\tau^{-\frac{d}{2}-(D-d)(1-\gamma)/2}F_{2}(\tau^ {-\frac{1}{2}}({\bf
y}    -{\bf x}),\tau^{-\frac{1}{2}+\frac{\gamma}{2}} ({\bf
Y}    -{\bf X}))
\end{equation}

 \section{A  projection to $D-d$ dimensions}

 The two-point function (18) has a different scaling behaviour
 in ${\bf x}$ and ${\bf X}$ directions. We  obtain
 a fixed scaling behaviour
  setting ${\bf x}    ={\bf y}=0    $. Then,  we have
  \begin{equation}
  \langle {\cal A}^{-1}(x,y)\rangle=  R  \vert
  {\bf X}-{\bf Y}\vert^{-D+2-\frac{\gamma}{1-\gamma}(d-2)}
  \end{equation}
  where $R$ is a positive constant.
   Hence, if  all the correlation  functions are scale invariant then
  \begin{equation}
  \Phi(0,{\bf X})\simeq
  \lambda^{\frac{D-2}{2}+\frac{\gamma}{1-\gamma}\frac{d-2}{2}}
  \Phi(0,\lambda{\bf X})
\end{equation}
where the equivalence means that both sides have the same
correlation functions.

In order to prove that $R$ is finite and not zero we need upper and
 lower  bounds for the Gaussian model (17).
 We show first that the
 bilinear form $(f_{j},\langle{\cal A}^{-1}\rangle f_{l})$
is finite and non-zero on a dense set of functions $f$. For this purpose we choose
\begin{displaymath}
f_{{\bf k}}({\bf X})= (2\pi a)^{-\frac{d}{2}}\exp(-\frac{a}{2}{\bf
X}^{2}+i{\bf k}{\bf X})
\end{displaymath}
Then, (we keep ${\bf x}\neq {\bf y}$ in order to show that the
model of sec.2 is non-trivial; for a scale invariant model of this
section ${\bf x}={\bf y}={\bf 0}$)
\begin{equation}
\begin{array}{l}
(f_{{\bf k}},\langle{\cal A}^{-1}\rangle f_{{\bf k}^{\prime}})=
(2\pi)^{-D+d}\int_{0}^{\infty}d\tau\tau^{-\frac{d}{2}}\int d{\bf P}
 \cr E[\delta \left(\tau^{-\frac{1}{2}}\left({\bf y}-{\bf
x}\right)-{\bf b}\left(1\right)\right) \cr
\exp\Big(-\frac{1}{2a}\left({\bf P}-{\bf k}\right)^{2}
-\frac{1}{2a}\left({\bf P}-{\bf k}^{\prime}\right)^{2} \cr
-\frac{1}{4}\tau^{2-2\gamma}\int_{0}^{1}P_{\mu}P_{\sigma}
P_{\nu}P_{\rho}D^{\mu\nu;\sigma\rho} \left ({\bf b}
\left(s\right)-{\bf b}\left(s^{\prime}\right)\Big)
dsds^{\prime}\right)]
\end{array}
\end{equation}
In our estimates  we apply Jensen inequalities in the form (for
real functions $A$ and $f$)
\begin{equation}
E[\exp A]\geq \exp E[A]
\end{equation}
and
\begin{equation}
 E[\exp\Big( -\int_{0}^{1} ds ds^{\prime}f(s,s^{\prime})\Big)]\leq
\int_{0}^{1} dsds^{\prime} E[\exp(-f(s,s^{\prime}))]
 \end{equation}
An upper bound can be obtained by means of the Jensen inequality
(23) expressed in the form
\begin{equation}
\begin{array}{l}
(f_{{\bf k}},\langle{\cal A}^{-1}\rangle f_{{\bf k}^{\prime}})
\cr
 \leq
2\int_{0}^{\infty}d\tau\int_{0}^{1}ds\int_{0}^{s}ds^{\prime}\int
d{\bf u}_{1}d{\bf u}_{2}d{\bf P}
 \cr
 \tau^{-\frac{d}{2}}\exp\left(-\frac{1}{2a}\left({\bf P}-{\bf k}\right)^{2}
-\frac{1}{2a}\left({\bf P}-{\bf k}^{\prime}\right)^{2}\right)
 p(s^{\prime},{\bf
u}_{1})p(s-s^{\prime},{\bf u}_{2}-{\bf u}_{1})
\cr
p\left(1-s,\tau^{-\frac{1}{2}}\left({\bf
y} -{\bf x}\right)-{\bf u}_{2}\right)
\exp\left(-\frac{\tau^{2-2\gamma}}{4}P_{\mu}P_{\sigma}
P_{\nu}P_{\rho}D^{\mu\nu;\sigma\rho}\left ({\bf u}_{1}-{\bf
u}_{2}\right)\right)
\end{array}
\end{equation}
where $p(s,{\bf u})=(2\pi s)^{-\frac{d}{2}}\exp(-{\bf u}^{2}/2s)$.
We can convince ourselves by means of explicit calculations
(using a proper change of variables) that the integral on the
r.h.s. of eq.(24) is finite. For  the lower bound it will be
useful to introduce the Brownian bridge  \cite{simon}
starting from ${\bf 0}$ and ending in ${\bf u}$ defined on
the time interval $[0,1]$
\begin{displaymath}
{\bf a}({\bf u},s)={\bf u}s+{\bf c}(s)
\end{displaymath}
where ${\bf c}$ is the Gaussian  process starting from ${\bf 0}$
and ending in  ${\bf 0}$ with mean equal zero and the covariance
\begin{displaymath}
E[c_{j}(s^{\prime})c_{k}(s)]=\delta_{jk}s^{\prime}(1-s)
\end{displaymath}
for $s^{\prime}\leq s$. Then,  the $\delta$  function in eq.(21)
determines the Brownian bridge  and the Jensen inequality (22)
takes the form

\begin{equation}
\begin{array}{l}
(f_{{\bf k}},\langle{\cal A}^{-1}\rangle f_{{\bf k}^{\prime}})
\geq (2\pi)^{-D+d}\int_{0}^{\infty}d\tau\tau^{-\frac{d}{2}}\int
d{\bf P}

\cr \exp\Big(-\frac{1}{2a}\left({\bf P}-{\bf k}\right)^{2}
-\frac{1}{2a}\left({\bf P}-{\bf k}^{\prime}\right)^{2} \cr
-\frac{1}{4}\tau^{2-2\gamma}\int_{0}^{1}P_{\mu}P_{\sigma}
P_{\nu}P_{\rho}E[D^{\mu\nu;\sigma\rho}\left ({\bf
a}\left(\tau^{-\frac{1}{2}}{\bf y}- \tau^{-\frac{1}{2}}{\bf x
},s\right) -{\bf a}\left(\tau^{-\frac{1}{2}}{\bf y}-
\tau^{-\frac{1}{2}}{\bf x},s^{\prime}\right)\Big)
dsds^{\prime}]\right)
\end{array}
\end{equation}
where the expectation value in  the exponential on the r.h.s. of
eq.(25) is equal to
\begin{equation}
\begin{array}{l}
\int d{\bf u}\int ds\int_{0}^{s}ds^{\prime} \left(2\pi
\omega\left(s,s^{\prime}\right)\right)^{-\frac{d}{2}} \exp\left(-
\frac{1}{2\omega(s,s^{\prime})}{\bf u}^{2}\right) \cr \vert {\bf
u}-\tau^{-\frac{1}{2}}s\left({\bf y}-{\bf x}\right)+
\tau^{-\frac{1}{2}}s^{\prime}\left({\bf y}-{\bf x}\right)
\vert^{-4\gamma}
\end{array}
\end{equation}
where $\omega(s,s^{\prime})=(s-s^{\prime})(1-s+s^{\prime})$. It
is finite if $\gamma<\frac{1}{2}$ (the form (16) of the graviton
two-point function is assumed).

 We  compute now higher order correlation functions
 in the Gaussian model
\begin{equation}
\begin{array}{l}
\langle \Phi(x)\Phi(x^{\prime})\Phi^{*}(y)\Phi^{*}(y^{\prime})\rangle

\cr

=\langle {\cal A}^{-1}\left(x,y\right) {\cal
A}^{-1}\left(x^{\prime},y^{\prime}\right)\rangle +(x\rightarrow
x^{\prime})
 \cr =(2\pi)^{-2D+2d}\int
d\tau_{1}d\tau_{2} \int d{\bf P}d{\bf P}^{\prime}
  \exp\left(i{\bf
P}\left({\bf Y}-{\bf X}\right)
 +i{\bf P}^{\prime}\left({\bf Y}^{\prime}-{\bf X}^{\prime}\right) \right)
 \cr
E[\delta \left({\bf y}-{\bf x}-{\bf
b}\left(\tau_{1}\right)\right) \delta\left({\bf
y}^{\prime}-{\bf x}^{\prime} -{\bf
b}^{\prime}\left(\tau_{2}\right) \right)
\cr \exp\Big( -\frac{1}{4}\int_{0}^{\tau_{1}}\int_{0}^{\tau_{1}}
P_{\mu}P_{\sigma} P_{\nu}P_{\rho}D^{\mu\nu;\sigma\rho}\left ({\bf
b}\left(s\right)-{\bf b}\left(s^{\prime}\right)\right)
dsds^{\prime} \cr -\frac{1}{4}\int_{0}^{\tau_{2}}
\int_{0}^{\tau_{2}} P_{\mu}^{\prime}P_{\sigma}^{\prime}
P_{\nu}^{\prime}P_{\rho}^{\prime}D^{\mu\nu;\sigma\rho}\left ({\bf
b}^{\prime}\left(s\right)- {\bf
b}^{\prime}\left(s^{\prime}\right)\right) dsds^{\prime} \cr
-\frac{1}{2}\int_{0}^{\tau_{1}}\int_{0}^{\tau_{2}} P_{\mu}P_{\nu}
P_{\rho}^{\prime}P_{\sigma}^{\prime}D^{\mu\nu;\sigma\rho}\left
({\bf x}-{\bf x}^{\prime}+{\bf b}\left(s\right)-{\bf
b}^{\prime}\left(s^{\prime}\right)\right)
dsds^{\prime}\Big)]     +(x\rightarrow x^{\prime})
\end{array}
\end{equation}
where $(x\rightarrow x^{\prime})$ means the same expression in which
$x$ is exchanged with $x^{\prime}$.
 The fourlinear form (27) calculated on   the basis $f$
reads
\begin{equation}
\begin{array}{l}
\langle \Phi(f_{{\bf k}_{1}})
\Phi(f_{{\bf k}_{3}})\Phi^{*}(f_{{\bf k}_{2}})\Phi^{*}(f_{{\bf k}_{4}}) \rangle

\cr
  =(2\pi)^{-2D+2d}\int
d\tau_{1}d\tau_{2} \int d{\bf P}d{\bf P}^{\prime}  E[\delta
\left({\bf y}-{\bf x}-{\bf b}\left(\tau_{1}\right)\right)
\delta\left({\bf y}^{\prime}-{\bf x}^{\prime} -{\bf
b}^{\prime}\left(\tau_{2}\right) \right) \cr \exp\left(
-\frac{1}{2a}({\bf P}-{\bf k}_{1})^{2} -\frac{1}{2a}({\bf P}-{\bf
k}_{2})^{2} -\frac{1}{2a}({\bf P}^{\prime}-{\bf k}_{3})^{2}
-\frac{1}{2a}({\bf P}^{\prime}-{\bf k}_{4})^{2} \right) \cr
\exp\Big( -\frac{1}{4}\int_{0}^{\tau_{1}}\int_{0}^{\tau_{1}}
P_{\mu}P_{\sigma} P_{\nu}P_{\rho}D^{\mu\nu;\sigma\rho}\left ({\bf
b}\left(s\right)-{\bf b}\left(s^{\prime}\right)\right)
dsds^{\prime} \cr -\frac{1}{4}\int_{0}^{\tau_{2}}
\int_{0}^{\tau_{2}} P_{\mu}^{\prime}P_{\sigma}^{\prime}
P_{\nu}^{\prime}P_{\rho}^{\prime}D^{\mu\nu;\sigma\rho}\left ({\bf
b}^{\prime}\left(s\right)- {\bf
b}^{\prime}\left(s^{\prime}\right)\right) dsds^{\prime} \cr
-\frac{1}{2}\int_{0}^{\tau_{1}}\int_{0}^{\tau_{2}} P_{\mu}P_{\nu}
P_{\rho}^{\prime}P_{\sigma}^{\prime}D^{\mu\nu;\sigma\rho}\left
({\bf x}-{\bf x}^{\prime}+{\bf b}\left(s\right)-{\bf
b}^{\prime}\left(s^{\prime}\right)\right)
dsds^{\prime}\Big)]+(1,2\rightarrow 3,4)
\end{array}
\end{equation}
where the last term means the same expression with exchanged wave
numbers. We introduce the spherical coordinates on the
$(\tau_{1},\tau_{2})$-plane $\tau_{1}=r\cos\theta$ and
$\tau_{2}=r\sin\theta$. Let us rescale the momenta ${\bf k}={\bf
p}\sqrt{r}$, ${\bf k}^{\prime} ={\bf p}^{\prime}\sqrt{r}$ , ${\bf
K}={\bf P}r^{\frac{1}{2}-\frac{\gamma}{2}}$  and
 ${\bf K}^{\prime}
={\bf P}^{\prime}r^{\frac{1}{2}-\frac{\gamma}{2}}$. Then, we can
see that the four-point function (27) takes the form
\begin{equation}
\begin{array}{l}
\langle
\Phi(x)\Phi(x^{\prime})\Phi^{*}(y)\Phi^{*}(y^{\prime})\rangle \cr
=\int d\theta drr r^{-d-(1-\gamma)(D-d)}F_{4}(\theta,
r^{-\frac{1}{2}}({\bf x}-{\bf y}), r^{-\frac{1}{2}}({\bf
x}^{\prime}-{\bf y}^{\prime}), \cr r^{-\frac{1}{2}}({\bf
x}^{\prime}-{\bf x}), r^{-\frac{1}{2}}({\bf
x}^{\prime}-{\bf y}), r^{-\frac{1}{2}}({\bf
y}^{\prime}-{\bf x}), \cr
r^{-\frac{1}{2}+\frac{\gamma}{2}} ({\bf X}-{\bf Y}),
r^{-\frac{1}{2}+\frac{\gamma}{2}} ({\bf X}^{\prime}-{\bf
Y}^{\prime}), r^{-\frac{1}{2}+\frac{\gamma}{2}} ({\bf
X}^{\prime}-{\bf Y}),
r^{-\frac{1}{2}+\frac{\gamma}{2}} ({\bf X}-{\bf
Y}^{\prime}))

\end{array}
\end{equation}
It follows just by scaling of coordinates (the $r$-integral
scales as twice the $\tau $-integral in eq.(18))
  that at ${\bf x}={\bf x}^{\prime}={\bf y}={\bf y}^{\prime}=0$
  the correlations are scale invariant with the same
  scaling dimension as in eq.(20).

 It is clear from eq.(28) that
 in the same way as we did it in eqs.(24)-(25)
 we can obtain finite upper
 and lower bounds on the correlation functions (28) by means of
 the Jensen inequalities.

We could continue with higher order correlation functions. Again
through an introduction of spherical coordinates in the
$(\tau_{1},...,\tau_{3})$ space  we can show that
\begin{equation}
\begin{array}{l}
\langle \Phi(x_{1}).....\Phi(x_{3}) \Phi^{*}(y_{1})....
\Phi^{*}(y_{3})\rangle
\end{array}
\end{equation}
scales with the same dimension as in eq.(20). The scaling of
higher order correlation functions is now evident. We introduce
the spherical coordinates for the $\tau$-integrals. The resulting
scaling is a consequence of the fact that the $\tau$-volume and
${\bf P}$ integrals have the scaling dimensions  proportional to
the order of the correlation function.

\section{ A  projection to $d$ dimensions}
There is still another option that we let all $X=Y=0$.
In such a case
\begin{equation}
\begin{array}{l}
\langle K_{\tau}({\bf x},{\bf y})\rangle=(2\pi)^{-D+d}\int d{\bf P}
 \cr E[\delta \left({\bf y}-{\bf
x}    -\sqrt{\tau}{\bf b}\left(1\right)\right)\cr
\exp\left(-\frac{1}{4}\tau^{2-2\gamma}\int_{0}^{1} P_{\mu}P_{\sigma}
P_{\nu}P_{\rho}D^{\mu\nu;\sigma\rho}\left ({\bf
b}\left(s\right)-{\bf b}\left(s^{\prime}\right)\right)
dsds^{\prime}\right)]
\end{array}
\end{equation}
 By
a scaling  of momenta we can bring the propagator of eq.(3) to
the form
\begin{equation}
\langle {\cal A}^{-1}(x,y)\rangle=\int_{0}^{\infty}d\tau
\tau^{-\frac{d}{2}-(D-d)(1-\gamma)/2}F_{2}(\tau^ {-\frac{1}{2}}({\bf
y}    -{\bf x}))
\end{equation}
 Hence
   \begin{equation}
  \langle {\cal A}^{-1}({\bf x},{\bf y})\rangle=  R  \vert
  {\bf x}-{\bf y}\vert^{-d+2-(D-d)(1-\gamma)}
  \end{equation}
  where $R$ is a positive constant.
   Hence, if  all the correlation  functions are scale invariant then
  \begin{equation}
  \Phi({\bf x},0)\simeq
  \lambda^{\frac{d-2}{2}+\frac{(D-d)(1-\gamma)}{2}}
  \Phi(\lambda{\bf x},0)
\end{equation}
We can prove all the inequalities of sec.3 in this model.
So, the upper bound for the two-point function reads
\begin{equation}
\begin{array}{l}
\vert\langle{\cal A}^{-1}({\bf x},{\bf y})\rangle\vert =
\cr
 \leq
2\int_{0}^{\infty}d\tau\int_{0}^{1}ds\int_{0}^{s}ds^{\prime}\int
d{\bf u}_{1}d{\bf u}_{2}d{\bf P}
 \tau^{-\frac{d}{2}}
 p(s^{\prime},{\bf
u}_{1})p(s-s^{\prime},{\bf u}_{2}-{\bf u}_{1})
\cr
p\left(1-s,\tau^{-\frac{1}{2}}\left({\bf
y} -{\bf x}\right)-{\bf u}_{2}\right)
\exp\left(-\frac{\tau^{2-2\gamma}}{4}P_{\mu}P_{\sigma}
P_{\nu}P_{\rho}D^{\mu\nu;\sigma\rho}\left ({\bf u}_{1}-{\bf
u}_{2}\right)\right)
\end{array}
\end{equation}
The lower bound takes the form

\begin{equation}
\begin{array}{l}
\vert \langle{\cal A}^{-1}({\bf x},{\bf y})\rangle \vert
\geq (2\pi)^{-D+d}\int_{0}^{\infty}d\tau\tau^{-\frac{d}{2}}\int
d{\bf P}

\cr \exp\Big(
-\frac{1}{4}\tau^{2-2\gamma}\int_{0}^{1}P_{\mu}P_{\sigma}
P_{\nu}P_{\rho}E[D^{\mu\nu;\sigma\rho}\left ({\bf
a}\left(\tau^{-\frac{1}{2}}{\bf y}- \tau^{-\frac{1}{2}}{\bf x
},s\right) -{\bf a}\left(\tau^{-\frac{1}{2}}{\bf y}-
\tau^{-\frac{1}{2}}{\bf x},s^{\prime}\right)\Big)
dsds^{\prime}]\right)
\end{array}
\end{equation}
where the expectation value in  the exponential on the r.h.s. of
eq.(36) is equal to
\begin{equation}
\begin{array}{l}
\int d{\bf u}\int
ds\int_{0}^{s}ds^{\prime}\left(2\pi\omega\left(s,s^{\prime}\right)\right)
^{-\frac{d}{2}} \exp\left(- \frac{1}{2\omega(s,s^{\prime})}{\bf
u}^{2}\right) \cr \vert {\bf u}-\tau^{-\frac{1}{2}}s\left({\bf
y}-{\bf x}\right)+ \tau^{-\frac{1}{2}}s^{\prime}\left({\bf
y}-{\bf x}\right) \vert^{-4\gamma}
\end{array}
\end{equation}
where $\omega(s,s^{\prime})=(s-s^{\prime})(1-s+s^{\prime})$. It
is finite if $\gamma<\frac{1}{2}$. The bounds (35)-(36) in fact
have the form
 \begin{equation}
 R_{1}\leq \vert {\bf x}-{\bf y}\vert ^{d-2+(D-d)(1-\gamma)}
 \langle {\cal A}^{-1}({\bf x},{\bf y})\rangle\leq R_{2}
 \end{equation}
 with certain positive $R_{1}$ and $R_{2}$.

 The inequalities (38) can be proved from the inequlities (35)-(36)
 just by rescaling of variables. It is more tedious to
 show that the constants $R_{1}$ and $R_{2}$ are finite
 and not zero (but the estimates reduce to finite
 dimensional integrals and are straightforward).

 We  can project  now to $R^{d}$ higher order correlation functions
\begin{equation}
\begin{array}{l}
\langle \Phi({\bf x})\Phi({\bf x}^{\prime})\Phi^{*}({\bf y})
\Phi^{*}({\bf y}^{\prime})\rangle

\cr

=\langle {\cal A}^{-1}\left({\bf x},{\bf y}\right) {\cal
A}^{-1}\left({\bf x}^{\prime},{\bf y}^{\prime}\right)\rangle +({\bf x}\rightarrow
{\bf x}^{\prime})
 \cr =(2\pi)^{-2D+2d}\int
d\tau_{1}d\tau_{2} \int d{\bf P}d{\bf P}^{\prime}
  \cr
E[\delta \left({\bf y}-{\bf x}-{\bf
b}\left(\tau_{1}\right)\right) \delta\left({\bf
y}^{\prime}-{\bf x}^{\prime} -{\bf
b}^{\prime}\left(\tau_{2}\right) \right)
\cr \exp\Big( -\frac{1}{4}\int_{0}^{\tau_{1}}\int_{0}^{\tau_{1}}
P_{\mu}P_{\sigma} P_{\nu}P_{\rho}D^{\mu\nu;\sigma\rho}\left ({\bf
b}\left(s\right)-{\bf b}\left(s^{\prime}\right)\right)
dsds^{\prime} \cr -\frac{1}{4}\int_{0}^{\tau_{2}}
\int_{0}^{\tau_{2}} P_{\mu}^{\prime}P_{\sigma}^{\prime}
P_{\nu}^{\prime}P_{\rho}^{\prime}D^{\mu\nu;\sigma\rho}\left ({\bf
b}^{\prime}\left(s\right)- {\bf
b}^{\prime}\left(s^{\prime}\right)\right) dsds^{\prime} \cr
-\frac{1}{2}\int_{0}^{\tau_{1}}\int_{0}^{\tau_{2}} P_{\mu}P_{\nu}
P_{\rho}^{\prime}P_{\sigma}^{\prime}D^{\mu\nu;\sigma\rho}\left
({\bf x}-{\bf x}^{\prime}+{\bf b}\left(s\right)-{\bf
b}^{\prime}\left(s^{\prime}\right)\right)
dsds^{\prime}\Big)]     +(x\rightarrow x^{\prime})
\end{array}
\end{equation}
where $(x\rightarrow x^{\prime})$ means the same expression in
which $x$ is exchanged with $x^{\prime}$. We introduce the
spherical coordinates on the $(\tau_{1},\tau_{2})$-plane
$\tau_{1}=r\cos\theta$, $\tau_{2}=r\sin\theta$ and we rescale the
momenta ${\bf k}={\bf p}\sqrt{r}$, ${\bf k}^{\prime} ={\bf
p}^{\prime}\sqrt{r}$ , ${\bf K}={\bf
P}r^{\frac{1}{2}-\frac{\gamma}{2}}$. Then, we can see that the
four-point function (39) takes the form
\begin{equation}
\begin{array}{l}
\langle
\Phi({\bf x})\Phi({\bf x}^{\prime})\Phi^{*}({\bf y})\Phi^{*}({\bf y}^{\prime})
\rangle \cr
=\int d\theta drr r^{-d-(1-\gamma)(D-d)}F_{4}(\theta,
r^{-\frac{1}{2}}({\bf x}-{\bf y}), r^{-\frac{1}{2}}({\bf
x}^{\prime}-{\bf y}^{\prime}), \cr r^{-\frac{1}{2}}({\bf
x}^{\prime}-{\bf x}), r^{-\frac{1}{2}}({\bf
x}^{\prime}-{\bf y}), r^{-\frac{1}{2}}({\bf
y}^{\prime}-{\bf x}))
\end{array}
\end{equation}
The upper bound now reads
\begin{equation}
\begin{array}{l}
\vert\langle \Phi({\bf x})\Phi({\bf x}^{\prime})\Phi^{*}({\bf y})
\Phi^{*}({\bf y}^{\prime})\rangle \vert\leq
\cr
  2(2\pi)^{-2D+2d}\int
d\tau_{1}d\tau_{2}\int_{0}^{1}ds\int_{0}^{s}ds^{\prime} \int d{\bf P}d{\bf P}^{\prime}  E[\delta
\left({\bf y}-{\bf x}-\sqrt{\tau_{1}}{\bf b}\left(1\right)\right)
\cr
\delta\left({\bf y}^{\prime}-{\bf x}^{\prime} -\sqrt{\tau_{2}}{\bf
b}^{\prime}\left(1\right) \right) \cr
\exp\Big( -\frac{\tau_{1}^{2}}{4}
P_{\mu}P_{\sigma} P_{\nu}P_{\rho}D^{\mu\nu;\sigma\rho}\left (\sqrt{\tau_{1}}{\bf
b}\left(s\right)-\sqrt{\tau_{1}}{\bf b}\left(s^{\prime}\right)\right)
 \cr -\frac{\tau_{2}^{2}}{4} P_{\mu}^{\prime}P_{\sigma}^{\prime}
P_{\nu}^{\prime}P_{\rho}^{\prime}D^{\mu\nu;\sigma\rho}\left (
\sqrt{\tau_{2}}{\bf
b}^{\prime}\left(s\right)- \sqrt{\tau_{2}}{\bf
b}^{\prime}\left(s^{\prime}\right)\right)  \cr
-\frac{\tau_{1}\tau_{2}}{2} P_{\mu}P_{\nu}
P_{\rho}^{\prime}P_{\sigma}^{\prime}D^{\mu\nu;\sigma\rho}\left
({\bf x}-{\bf x}^{\prime}+\sqrt{\tau_{1}}{\bf b}\left(s\right)-
\sqrt{\tau_{2}}{\bf
b}^{\prime}\left(s^{\prime}\right)\right)
\Big)]+({\bf x}\rightarrow {\bf x}^{\prime})
\end{array}
\end{equation}
The expectation value (41) can be expressed by the transition
functions (as usual for the Wiener process). The bound is scale
invariant and the scale invariant function on the r.h.s. could be
calculated explicitly. The lower bound takes the form
\begin{equation}
\begin{array}{l}
\langle \Phi({\bf x})\Phi({\bf x}^{\prime})\Phi^{*}({\bf y})
\Phi^{*}({\bf y}^{\prime})\rangle

\cr
\geq  (2\pi)^{-2D+2d}\int
d\tau_{1}d\tau_{2} \int d{\bf P}d{\bf P}^{\prime}  \cr

\exp\Big( -E[\frac{1}{4}\int_{0}^{\tau_{1}}\int_{0}^{\tau_{1}}
P_{\mu}P_{\sigma} P_{\nu}P_{\rho}D^{\mu\nu;\sigma\rho}\left ({\bf
a}\left(s\right)-{\bf a}\left(s^{\prime}\right)\right)
dsds^{\prime} \cr \frac{1}{4}\int_{0}^{\tau_{2}}
\int_{0}^{\tau_{2}} P_{\mu}^{\prime}P_{\sigma}^{\prime}
P_{\nu}^{\prime}P_{\rho}^{\prime}D^{\mu\nu;\sigma\rho}\left ({\bf
a}^{\prime}\left(s\right)- {\bf
a}^{\prime}\left(s^{\prime}\right)\right) dsds^{\prime} \cr
+\frac{1}{2}\int_{0}^{\tau_{1}}\int_{0}^{\tau_{2}} P_{\mu}P_{\nu}
P_{\rho}^{\prime}P_{\sigma}^{\prime}D^{\mu\nu;\sigma\rho}\left
({\bf a}\left(s\right)-{\bf
a}^{\prime}\left(s^{\prime}\right)\right)  dsds^{\prime}]\Big)
+({\bf x}\rightarrow {\bf x}^{\prime})
\end{array}
\end{equation}
Here
\begin{displaymath}
{\bf a}(s)={\bf x}+({\bf y}-{\bf x})\frac{s}{\tau_{1}}+
\sqrt{\tau_{1}}{\bf c}(\frac{s}{\tau_{1}})
\end{displaymath}
 and
\begin{displaymath}
{\bf a}^{\prime}(s)={\bf x}^{\prime}+({\bf y}^{\prime}-{\bf x}^{\prime})\frac{s}{\tau_{2}}+
\sqrt{\tau_{2}}{\bf c}^{\prime}(\frac{s}{\tau_{2}})
\end{displaymath}
In the exponential of the formula (42) we have the expectation
over three Gaussian processes. The first is with the mean $({\bf
y}-{\bf x})(s-s^{\prime})/\tau_{1} $ and the covariance
$(s-s^{\prime})(\tau_{1}-s+s^{\prime})/\tau_{1}$, the second has
the mean
 $({\bf y}^{\prime}-{\bf x}^{\prime})(s-s^{\prime})/\tau_{2} $
and the covariance $(s-s^{\prime})(\tau_{2}-s+s^{\prime})/\tau_{2}$ ,
the third has the mean
${\bf x}-{\bf x}^{\prime}+({\bf y}-{\bf x})s/\tau_{1}
 -({\bf y}^{\prime}-{\bf x}^{\prime})
s^{\prime}/\tau_{2} $
 and the covariance
$s(\tau_{1}-s)/\tau_{1}+s^{\prime}(\tau_{2}-s^{\prime})/\tau_{2}$.
 The lower bound can be explicitly calculated and is given
 by a scale invariant function.
 It is clear how to calculate the higher order scale
 invariant functions and their scale invariant lower
 and upper bounds.

 \section{Discussion}
 The model  discussed in sec.3 is invariant under  Euclidean rotations in
$D-d$ dimensions.  Euclidean fields with the Osterwalder-Schrader
positivity cannot be more regular than the free field (this
follows from the K\"allen-Lehman representation).
 In $D$ dimensions the short distance  behaviour
of the correlation functions (18) is more regular than the one of
the free fields. However, after setting all ${\bf x}=0$ the
behaviour is more singular than the canonical one in $D-d$
dimensions.
    We can
suggest a lattice model
 whose formal continuum limit coincides with our scale
invariant Euclidean field theory. The simplest possibility is to
take the conformally flat metric placed on a sublattice just
between the lattice sites of the scalar field (as the gauge
fields in ref.\cite{ref1}). It seems that the Gaussian model of
sec.3 mixing the ferromagnetic and antiferromagnetic couplings
would fail to be reflection positive (in any case this would not
be easy to prove, see ref.\cite{ref2}). The model (11) with the
metric which is a square of a Gaussian field can be reflection
positive (it is scale invariant with correlation functions
expressed by the characteristic function ${\cal J}(h)$ (13)). The
continuum limit and the subsequent analytic continuation to
Minkowski space would give a model of relativistic quantum field
theory satisfying all Wightman axioms. The Wick square of a
Gaussian field is an example of an infinitely divisible field
\cite{bau}. An infinitely divisible field can take non-negative
values. Its characteristic function has an explicit integral
representation. Such a field can be a good candidate for a random
metric.

More interesting are the models in sec.4. The lattice version of
the Lagrangian
\begin{displaymath}
L=g^{\mu\nu}({\bf x})\frac{\partial}{\partial X^{\mu}}\Phi
\frac{\partial}{\partial X^{\nu}}\Phi^{*} +\nabla_{\bf x}\Phi
\nabla_{\bf x}\Phi^{*}
\end{displaymath}
will have the form  $-L=F+\theta F+M\theta M$ where $\theta $ is
the reflection in the plane perpendicular to one of the
coordinates  ${\bf x}$ (which will be chosen as time). This
representation holds true because the random metric does not mix
the temporal coordinates in $\nabla_{\bf x}\Phi\nabla_{\bf x}\Phi^{*}
$. Then, the reflection positivity results (see \cite{ref2} ). In
the lattice approximation we have to replace the (formal) Gaussian
measure with a negative definite covariance (16) by a
non-Gaussian measure on the metrics which has a formal Gaussian
limit (e.g., replacing $\frac{1}{2}x^{2}$ by $1-\cos x$). There
remains to be proven that such a lattice approximation is
convergent to the continuum.

\end{document}